\documentclass[11pt,aps,prd,preprint,superscriptaddress,showpacs]{revtex4}
\pdfoutput=1
\usepackage{revsymb}
\usepackage{amssymb}
\usepackage{amsmath}
\usepackage{amsfonts}
\usepackage{placeins}
\usepackage[pdftex]{graphicx}
\usepackage{subfigure}
\DeclareGraphicsExtensions{.pdf,.png,.jpg} 

\newcommand{\be}{\begin{equation}}
\newcommand{\ee}{\end{equation}}
\newcommand{\ben}{\begin{eqnarray}}
\newcommand{\een}{\end{eqnarray}}
\newcommand{\n}{\label}

\begin{document}
\title{CMB and matter power spectra with non-linear dark-sector interactions}
\author{R.F. vom Marttens\footnote{E-mail:rodrigovonmarttens@gmail.com}}
\affiliation{Universidade Federal do Esp\'{\i}rito Santo,
Departamento
de F\'{\i}sica\\
Av. Fernando Ferrari, 514, Campus de Goiabeiras, CEP 29075-910,
Vit\'oria, Esp\'{\i}rito Santo, Brazil}
\author{L. Casarini \footnote{E-mail:casarini.astro@gmail.com}}
\affiliation{Universidade Federal do Esp\'{\i}rito Santo,
Departamento
de F\'{\i}sica\\
Av. Fernando Ferrari, 514, Campus de Goiabeiras, CEP 29075-910,
Vit\'oria, Esp\'{\i}rito Santo, Brazil}
\affiliation{Institute of Theoretical Astrophysics, University of Oslo, 0315 Oslo, Norway}
\author{W.S. Hip\'{o}lito-Ricaldi\footnote{E-mail:wiliam.ricaldi@ufes.br}}
\affiliation{Universidade Federal do Esp\'{\i}rito Santo, Departamento de Ci\^encias  Naturais,\\
Rodovia BR 101 Norte, km. 60, CEP 29932-540,
S\~ao Mateus, Esp\'{\i}rito Santo, Brazil}
\author{W. Zimdahl\footnote{E-mail:winfried.zimdahl@pq.cnpq.br}}
\affiliation{Universidade Federal do Esp\'{\i}rito Santo,
Departamento
de F\'{\i}sica\\
Av. Fernando Ferrari, 514, Campus de Goiabeiras, CEP 29075-910,
Vit\'oria, Esp\'{\i}rito Santo, Brazil}

\date{\today}

\begin{abstract}
An interaction between dark matter and dark energy, proportional to the product of their energy densities, results in a scaling behavior of the ratio of these densities with respect to the scale factor of the Robertson-Walker metric. This gives rise to a class of cosmological models which deviate from the standard model in an analytically tractable way. In particular, it becomes possible to quantify the role of potential dark-energy perturbations.
 We investigate the impact of this interaction on the structure
formation process.  Using the (modified) CAMB code we obtain the CMB spectrum as well as the linear matter power spectrum. It is shown that the strong degeneracy in the parameter space present in the background analysis
is considerably reduced by considering \textit{Planck} data. Our analysis is compatible with the $\Lambda$CDM model at the $2\sigma$ confidence level with a slightly preferred
direction of the energy flow from dark matter to dark energy.
\end{abstract}


\maketitle

\section{Introduction}

According to the currently most favored cosmological model, the $\Lambda$CDM model
($\Lambda$ denotes the cosmological constant, CDM stands for cold dark matter), our observable
Universe is geometrically flat and ``normal", i.e. baryonic, matter is only responsible for about 4-5$\%$ of its content. Most of the Universe is composed of two exotic ``fluids": dark matter (DM) and dark energy (DE).
Neither of these exotic components can be observed directly so far, but there are arguments that support their existence.
DM explains galaxy rotation curves and plays a crucial role in cosmic structure formation.
DE is seen as an effective fluid with negative pressure, possibly associated with the quantum vacuum. This ``fluid" is able to account for the accelerated expansion of the scale factor of the Robertson-Walker metric of the standard model.
Even being a model with observational success, the $\Lambda$CDM model leaves open  the physical nature of these dark components. Moreover, there are a number of tensions which still await clarification (see, e.g., \cite{buchert15}).
There is therefore ongoing interest in discussing and probing alternative approaches which differ in one or more aspects from the $\Lambda$CDM dynamics.
One line of research does no longer regard
 DM and DE to be independent components with separate energy-momentum conservation  but investigates the consequences of a more complex structure of the dark sector, modeled by an interaction between its principal constituents.
Admitting the possibility of a non-gravitational coupling between DM and DE is, of course, the more general case which gives rise to a richer cosmological dynamics.
In particular, this generalization implies the existence  of DE perturbations, even for cases of an equation-of-state (EoS) parameter $w_{\Lambda} = -1$, sometimes called decaying vacuum models \cite{oesertaha,oesertaha87,berto,Freese,WangMeng,WangGongAbd,Borges}.
It has been pointed out that ignoring DE perturbations may result in unreliable conclusions concerning the
interpretation of observational data \cite{park}.
For further studies about clustering DE see, e.g., \cite{gawelaref,zhumberto,basilakos,duniya}.
Ignoring a potential interaction between both dark components altogether, may result in an incorrect interpretation of cosmological data \cite{dascorasaniti}.
Many models have been established by now which consider different types of  interactions \cite{wetterich,andrew,amendola,ZPC,CJPZ,CaiWang}.
Some more recent studies include \cite{hewang,royinstab,gawela,royobs,verde,salvatelli,
clemson,LiZhangZhang,faraoni,alonso,valerio,szyd,odder,zhuk,wangAbd1,wangAbd2,FengZhang}.
 All of them are phenomenological since the physical nature of the dark sector largely remains a matter of speculation.
In most cases interactions are assumed to be linear in the sense that the coupling terms in the energy-momentum balances of the components are proportional either to the DM density or to the DE density or to a linear combination of both densities (see, e.g., \cite{verde,wangAbd1,wangAbd2}).
There exists a large body of literature which studies such models, usually resulting in  limits on the type or the strength of the interaction.
Now, from a physical point of view it seems more natural to prefer an interaction that depends on the product of the abundances of the individual components, as, e.g., in chemical reactions.
It was shown in \cite{hewang} that a  coupling proportional to the product of the densities of DM and DE is also observationally favored over linear models.
Further recent studies of non-linearly interacting DE models  are \cite{zhuk} and \cite{FengZhang}.
Systems with non-linear interactions  do not allow, in most cases, an analytic treatment, not even for the homogeneous and isotropic background.
Here we consider the special case of a non-linear interaction for which there exists an analytic background solution.
While the existence of such analytic solution is important in itself, we emphasize its usefulness in setting up
the system of equations for the perturbation dynamics about this background.
This solution will be characterized by a single additional parameter the value of which serves to quantify and to restrict potential differences from the
$\Lambda$CDM model in a simple and transparent way.
Using a suitable modification of the CAMB code \cite{camb} we focus on the implications of the coupling on the anisotropy spectrum of the cosmic microwave background (CMB) and on the (total)  matter power spectrum.
This complements and extends previous work on the status of the non-linear model using large-scale-structure data \cite{alonso}.
In this context we also point out that the simple use of the position of the first acoustic peak in the CMB anisotropy spectrum, something which was occasionally done in the literature, is not a reliable criterion in assessing competing cosmological models.

The structure of the paper is as follows. In Sec.~\ref{gencos} we present the basic set of fluid dynamical equations both for the background and for the first-order perturbation dynamics of a multi-component cosmic medium. This system is applied to our model with non-linear interactions in the dark sector in Sec.~\ref{scaling}.
The subsequent Sec.~\ref{numeric} is devoted to the numerical and statistical analysis of our model.
Sec.~\ref{discussion} summarizes and discusses our results.

\section{Cosmology with interaction in the dark sector}
\label{gencos}

\subsection{General equations}

We consider a spatially flat universe with a perturbed Robertson-Walker (RW) metric which reads, up to first order in the perturbations and in the synchronous gauge,
\begin{equation}
ds^{2}=a^{2}\left[-d\tau^{2}+\left(\delta_{ij}+h_{ij}\right)dx^{i}dx^{j}\right].
\label{metric}
\end{equation}
Here, $a$ is the scale factor and $\tau$ is the conformal time.
Restricting ourselves to scalar perturbations, the first-order quantity
$h_{ij}$ has two scalars degrees of freedom $h$ and $\eta$ according to  \cite{ma&bertschinger},
\begin{equation}
h_{ij}\left(\tau,\vec{x}\right) =\int \left\lbrace e^{-i\vec{k}\cdot\vec{x}}\left[\hat{k}_{i}\hat{k}_{j}h\left(\tau,\vec{k}\right)+
\left(\hat{k}_{i}\hat{k}_{j}-\dfrac{1}{3}\delta_{ij}\right)6\eta\left(\tau,\vec{k}\right)\right]\right\rbrace d^{3}k.
\label{hij}
\end{equation}

We assume that the cosmic substratum is composed of a set of components that may or may not interact  with each other.  Each component, characterized by a subindex  $x$, will be treated as a perfect fluid with an
energy-momentum tensor
\begin{equation}
T^{\mu\nu}_{x}=\rho_{x}u^{\mu}_{x}u^{\nu}_{x}+p_{x}h^{\mu\nu}_{x},
\label{emtensor}
\end{equation}
where $u^{\mu}_{x}$ is the $x$-component's four-velocity and $h^{\mu\nu}_{x}=u^{\mu}_{x}u^{\nu}_{x}+g^{\mu\nu}$ is a projection tensor
orthogonal to the four-velocity. $\rho_{x}$ and $p_{x}$ are energy density and pressure, respectively, of component $x$.
Quantities without subindex $x$ will refer to the total fluid.

Like the metric tensor, the other dynamical quantities can be split into a sum of a zeroth-order term
(denoted by a bar) and a first-order term (denoted by a hat). Since zeroth-order quantities live in a homogeneous and isotropic
background, the dynamical quantities are
\begin{eqnarray}
\left\{ \begin{array}{ll}\rho_{x}\left(\tau,\vec{x}\right)=\bar{\rho}_{x}\left(\tau\right)+\hat{\rho}_{x}\left(\tau,\vec{x}\right);\\ \\
p_{x}\left(\tau,\vec{x}\right)=\bar{p}_{x}\left(\tau\right)+\hat{p}_{x}\left(\tau,\vec{x}\right);\\ \\u^{\mu}_{x}=\dfrac{1}{a}
\left(1,\partial^{i}\hat{v}_{x}\right). \end{array} \right.
\label{quantities}
\end{eqnarray}
In the last relation, the spatial perturbation of the four-velocity was written as a divergency of a scalar function, known as peculiar velocity
potential.

The total energy-momentum tensor is the sum of all the individual energy-momentum tensors. If a certain component interacts with one or more of the remaining components,
its energy-momentum balance is affected by an interaction term $Q^{\mu}_{x}$,
\begin{equation}
\nabla_{\mu}T^{\mu\nu}_{x}=Q^{\nu}_{x}.
\label{cderivative}
\end{equation}
Since we assume that gravity is described by general relativity (GR), the total energy-momentum tensor must  be conserved, i.e, the interaction terms
must satisfy the relation
\begin{equation}
\sum_{x}Q^{\mu}_{x}=0.
\label{bianchi}
\end{equation}
The interaction term is split in a similar way as the quantities in
equation (\ref{quantities}),
\begin{equation}
Q^{\mu}_{x}\left(\tau,\vec{x}\right)=\bar{Q}^{\mu}_{x}\left(\tau\right)+\hat{Q}^{\mu}_{x}\left(\tau,\vec{x}\right).
\label{splitq}
\end{equation}
Following \cite{kodama}, is convenient to decompose this interaction $Q^{\mu}$ into two terms: a term parallel to the four-velocity, and a term
orthogonal to the four-velocity,
\begin{equation}
Q^{\mu}_{x}=Q_{x}u^{\mu}_{x}+\hat{F}^{\mu}_{x},\quad\quad\mbox{where}\quad\quad \hat{F}^{\mu}_{x}u_{x\mu}=0.
\label{qu}
\end{equation}
The temporal component of the interaction term is associated to the energy transfer, and its spatial component to the transfer of momentum. Spatial homogeneity implies that the spatial component of the background interaction term must be identically zero.

Under these conditions the background dynamics is described by Friedmann's
equations,
\begin{eqnarray}
&\mathcal{H}^{2}=\dfrac{8\pi Ga^{2}}{3}\rho, \label{friedmann1}\\
&\mathcal{H}^{\prime}=-\dfrac{4\pi Ga^{2}}{3}\left(\rho+3p\right),\label{friedmann2}
\end{eqnarray}
while the first-order perturbation equations are
\begin{eqnarray}
&k^{2}\eta-\dfrac{1}{2}\mathcal{H}h^{\prime}=-4\pi Ga^{2}\hat{\rho},\label{peinstein1}\\
&\eta^{\prime}=-4\pi Ga^{2}\left(\bar{\rho}+\bar{p}\right)v,\label{peinstein2}\\
&h^{\prime\prime}+2\mathcal{H}h^{\prime}-2k^{2}\eta=-8\pi Ga^{2}\hat{p},\label{peinstein3}\\
&h^{\prime\prime}+6\eta^{\prime\prime}+2\mathcal{H}\left(h^{\prime}+6\eta^{\prime}\right)-2k^{2}\eta=-24\pi Ga^{2}\left(\rho+p\right)\sigma,\label{peinstein4}
\end{eqnarray}
where the prime denotes the derivative with respect to the conformal time, $\mathcal{H}\equiv \frac{a^{\prime}}{a}$ is the Hubble parameter with
respect to the conformal time and $\sigma$ is a quantity that can be associated with the anisotropic stress \cite{ma&bertschinger}.

The energy balance for each of the components can be obtained by the projection of the covariant derivative of the energy-momentum tensor on the four-velocity. In
zeroth order we assume the rest frames of the components to coincide.  For the zeroth-order energy balances  we have
\begin{equation}
\bar{\rho}^{\prime}_{x}+3\mathcal{H}\left(1+w_{x}\right)\bar{\rho}_{x}=a\bar{Q}_{x},
\label{bgenergy1}
\end{equation}
where $w_{x}$ is the background equation-of-state (EoS) parameter
\begin{equation}
w_{x}\equiv\dfrac{\bar{p}_{x}}{\bar{\rho}_{x}}.
\label{w}
\end{equation}
The first-order equation is
\begin{equation}
\hat{\rho}^{\prime}_{x}+3\mathcal{H}\left(\hat{\rho}_{x}+\hat{p}_{x}\right)
-\left(\bar{\rho}_{x}+\bar{p}_{x}\right)\left(k^{2}v_{x}+\dfrac{h^{\prime}}{2}\right)=a\hat{Q}_{x}.
\label{energygeneral}
\end{equation}
The momentum conservation is obtained by the projection of the covariant derivative orthogonal to the four-velocity. As already mentioned, the background momentum balance is identically zero, thus, the momentum contributes only in first order. It reads
\begin{equation}
\left[\left(\bar{\rho}_{x}+\bar{p}_{x}\right)v_{x}\right]^{\prime}+4\mathcal{H}\left(\bar{\rho}_{x}+\bar{p}_{x}\right)
v_{x}+\hat{p}_{x}=a\left(\bar{Q}_{x}v_{x}+\hat{f}_{x}\right).
\label{momentumgeneral}
\end{equation}
Here, the scalar function $\hat{f}_{x}$ was introduced via,
\begin{equation}
\hat{F}^{i}_{x}=\dfrac{1}{a}\partial^{i}\hat{f}_{x}.
\end{equation}
The adiabatic sound speed is defined by
\begin{equation}
c_{(a)x}^{2}=\dfrac{\bar{p}^{\prime}_{x}}{\bar{\rho}^{\prime}_{x}}=w_{x}+\dfrac{w^{\prime}_{x}\bar{\rho}_{x}}{\bar{\rho}^
{\prime}_{x}}.
\label{csadiabatic}
\end{equation}
To characterize the propagation of perturbations we introduce the comoving sound speed
\begin{equation}
c_{(s)x}^{2}=\dfrac{\hat{p}^{(c)}_{x}}{\hat{\rho}^{(c)}_{x}},
\label{cscomoving}
\end{equation}
where $\hat{p}^{(c)}_{x}$ and $\hat{\rho}^{(c)}_{x}$ are the gauge-invariant comoving perturbations of pressure and energy density, respectively:
\begin{eqnarray}
&\hat{p}^{(c)}_{x}=\hat{p}_{x}+\bar{p}^{\prime}_{x}v_{x}, \label{pcomoving}\\
&\hat{\rho}^{(c)}_{x}=\hat{\rho}_{x}+\bar{\rho}^{\prime}_{x}v_{x}. \label{rhocomoving}
\end{eqnarray}
An alternative way to write the pressure perturbations is
\begin{equation}
\hat{p}_{x}=c_{(s)x}^{2}\hat{\rho}_{x}+\left(c_{(s)x}^{2}-c_{(a)x}^{2}\right)\bar{\rho}^{\prime}_{x}v_{x}.
\label{hatp}
\end{equation}
The density contrast of component x will be described by
\begin{equation}
\delta_{x}\equiv\dfrac{\hat{\rho}_{x}}{\bar{\rho}_{x}}\quad\Rightarrow\quad\delta^{\prime}_{x}=\dfrac{\hat{\rho}^{\prime}_{x}}
{\bar{\rho}_{x}}-\dfrac{\bar{\rho}^{\prime}_{x}}{\bar{\rho}_{x}}\delta_{x}.
\label{deltageral}
\end{equation}

Now, using the definitions (\ref{w}), (\ref{csadiabatic}), (\ref{cscomoving}) and (\ref{deltageral}), the first-order
energy balance (\ref{energygeneral}) in the $k$-space becomes
\begin{eqnarray}
\delta^{\prime}_{x}&+&3\mathcal{H}\left(c^{2}_{(s)x}-w_{x}\right)\delta_{x}
\nonumber\\
&-&9\mathcal{H}^{2}\left(1+w_{x}\right)\left(c^{2}_{(s)x}-
w_{x}\right)v_{x}
-\left(1+w_{x}\right)\left(k^{2}v_{x}-\dfrac{h^{\prime}}{2}\right)
+3\mathcal{H}w^{\prime}_{x}v_{x}
\nonumber\\
&& \qquad =\dfrac{a}{\bar{\rho}_{x}}\hat{Q}_{x} -\dfrac{a\bar{Q}_{x}}{\bar{\rho}_{x}}\left[\delta_{x}
+3\mathcal{H}\left(c^{2}_{(s)x}-w_{x}\right)v_{x}\right],
\label{kenergy}
\end{eqnarray}
while the momentum balance (\ref{momentumgeneral}) yields,
\begin{eqnarray}
v^{\prime}_{x}\left(1+w_{x}\right)+\mathcal{H}\left(1-3c^{2}_{(s)x}\right)\left(1+w_{x}\right)v_{x}
+c^{2}_{(s)x}\delta_{x}\qquad\qquad&&
\nonumber\\
=\dfrac{a\bar{Q}_{x}}{\bar{\rho}_{x}}\left[v-\left(1+c^{2}_{(s)x}\right)v_{x}\right]-\dfrac{a}
{\bar{\rho}_{x}}\hat{f}_{x}.&&
\label{kmomentum}
\end{eqnarray}
These equations hold separately for each component. We shall model now the content of the
Universe as a four-component fluid, consisting of  radiation (index $r$), baryonic matter (index $b$),
CDM (index $c$) and DE (index $\Lambda$).
We assume that there is an interaction between the dark components with
(cf.~(\ref{bianchi}))
\begin{equation}
Q^{\mu}=Q^{\mu}_{\Lambda}=-Q^{\mu}_{c},
\label{q1dark}
\end{equation}
while baryons and radiation  behave in the same way as in the $\Lambda$CDM model.
Neither baryons nor photons couple directly to DE and DM. They interact with each other via Thomson scattering  before recombination.
Thus, the equations for these two  components are the well-established Boltzmann equations of \cite{ma&bertschinger}.
For the dark sector we use
the fluid eqs. (\ref{kenergy}) and (\ref{kmomentum})
since a microscopic description of DE and DM with corresponding Boltzmann equations is not yet available.

\subsection{Dark sector perturbations}

Now we apply equations (\ref{kenergy}) and (\ref{kmomentum}) to the CDM and DE components.
For a constant EoS parameter of the DE component this parameter coincides with the adiabatic sound speed.
Under this condition the DE energy balance becomes
\begin{eqnarray}
\delta^{\prime}_{\Lambda}&+&3\mathcal{H}\left(c^{2}_{(s)\Lambda}-w_{\Lambda}\right)\delta_{\Lambda}
\nonumber\\
&-&9\mathcal{H}^{2}\left(1+w_{\Lambda}
\right)\left(c^{2}_{(s)\Lambda}-w_{\Lambda}\right)v_{\Lambda} - \left(1+w_{\Lambda}\right)\left(k^{2}v_{\Lambda}-\dfrac{h^{\prime}}{2}\right)
\nonumber\\
&&=\dfrac{a}{\bar{\rho}_{\Lambda}}\hat{Q}
-\dfrac{a\bar{Q}}{\bar{\rho}_{\Lambda}}\left[\delta_{\Lambda}+3\mathcal{H}\left(c^{2}_{(s)\Lambda}-w_{\Lambda}\right)
v_{\Lambda}\right]
\label{deenergypert}
\end{eqnarray}
and the momentum balance reduces to
\begin{eqnarray}
v^{\prime}_{\Lambda}\left(1+w_{\Lambda}\right)+\mathcal{H}\left(1-3c^{2}_{(s)\Lambda}\right)\left(1+w_{\Lambda}\right)v_{\Lambda}
+c^{2}_{(s)\Lambda}\delta_{\Lambda}\quad \qquad&&\nonumber\\
=\dfrac{a\bar{Q}}{\bar{\rho}_{\Lambda}}\left[v-\left(1+c^{2}_{(s)\Lambda}\right)v_{\Lambda}\right]-\dfrac{a}
{\bar{\rho}_{\Lambda}}\hat{f}.&&
\label{demomentumpert}
\end{eqnarray}

The analysis of the DE dynamics has to be performed separately for the cases
$w_{\Lambda}=-1$ and  $w_{\Lambda}
\neq -1$.
Equation (\ref{demomentumpert}) shows that for $w_{\Lambda}=-1$ the DE peculiar velocity is not a dynamic variable.  Then, this equation can be
used to determine $\hat{f}$ explicitly in a model-independent way,
\begin{equation}
\hat{f}=\dfrac{c^{2}_{(s)\Lambda}\rho_{\Lambda}}{a}\delta_{\Lambda}-\bar{Q}\hat{v}.
\label{hatf}
\end{equation}
This result will be used later in the CDM equation. For $w_{\Lambda}=-1$ the DE energy balance takes the form
\begin{equation}
\delta^{\prime}_{\Lambda}+3\mathcal{H}\left(c^{2}_{(s)\Lambda}+1\right)\delta_{\Lambda}
=\dfrac{a}{\bar{\rho}_{\Lambda}}\hat{Q}
-\dfrac{a\bar{Q}}{\bar{\rho}_{\Lambda}}\left[\delta_{\Lambda}+3\mathcal{H}\left(c^{2}_{(s)\Lambda}
+1\right)v_{\Lambda}\right].
\label{deenergypert1}
\end{equation}
In the second case,$w_{\Lambda}\neq -1$, the DE peculiar velocity is dynamically relevant and equation (\ref{demomentumpert}) must be solved to determine the
temporal evolution of $v_{\Lambda}$. There is no way to determine the perturbation of the spatial interaction term as in the previous case. Therefore, we restrict ourselves to
$\hat{f}=0$. The energy balance is again given by equation (\ref{deenergypert}), while the momentum balance simplifies to
\begin{eqnarray}
v^{\prime}_{\Lambda}\left(1+w_{\Lambda}\right)+\mathcal{H}\left(1-3c^{2}_{(s)\Lambda}\right)\left(1+w_{\Lambda}\right)v_{\Lambda}
+c^{2}_{(s)\Lambda}\delta_{\Lambda}\qquad\qquad &&\nonumber\\
=\dfrac{a\bar{Q}}{\bar{\rho}_{\Lambda}}\left[v-\left(1+c^{2}_{(s)\Lambda}\right)v_{\Lambda}\right].&&
\label{demomentumpert1}
\end{eqnarray}

The CDM component is characterized by $w_{c}=0$ with negligible pressure perturbations. Using the definitions (\ref{csadiabatic}),
(\ref{pcomoving}) and (\ref{cscomoving}) one concludes that $c^{2}_{(a)c}=\hat{c}^{2}_{(s)c}=0$. Its perturbative energy balance is
\begin{eqnarray}
&\delta^{\prime}_{c}-k^{2}v_{c}+\dfrac{h^{\prime}}{2}=-\dfrac{a}{\bar{\rho}_{c}}\hat{Q}+
\dfrac{a\bar{Q}}{\bar{\rho}_{c}}\delta_{c}.
\label{cdmenergypert}
\end{eqnarray}
The perturbative CDM momentum balance is affected by the $\hat{f}$ term. For $w_{\Lambda}=-1$ use of
(\ref{hatf}) provides us with
\begin{equation}
v^{\prime}_{c}+\mathcal{H}v_{c}
=-\dfrac{a\bar{Q}}{\bar{\rho}_{c}}\left(v-v_{c}\right)+\dfrac{a}{\bar{\rho}_{c}}\hat{f},
\label{cdmmomentumpert}
\end{equation}
where $\hat{f}$ is given by equation (\ref{hatf}).

In the second case with $\hat{f} =0$ the perturbative momentum balance reduces to
\begin{equation}
v^{\prime}_{c}+\mathcal{H}v_{c}=-\dfrac{a\bar{Q}}{\bar{\rho}_{c}}\left(v-v_{c}\right).
\label{cdmmomentumpert1}
\end{equation}
Our relations so far are valid for  any cosmological model consisting of an effective DE fluid which may interact with  pressureless DM as well as of baryons and radiation, both the latter being treated as in the standard model \cite{ma&bertschinger}.
In  the following we shall consider a specific configuration  which is analytically solvable in the homogeneous  and  isotropic background and can be understood as the consequence of a coupling between DM and DE that is proportional to the product of the energy densities of both dark components.

\section{Scaling cosmology model}
\label{scaling}

Our model of interest in this paper relies on a scaling behavior of the ratio of the energy densities of DM and DE. It can be demonstrated that it is this dynamics which is generated by an interaction term proportional to the product of the densities of DM and DE.
Following \cite{alonso}, we start by introducing a covariant length scale $l$ by \cite{ellis}
\begin{equation}
\dfrac{\dot{l}}{l}=\dfrac{1}{3}\Theta, \qquad \dot{l} \equiv l_{,\alpha}u^{\alpha}.
\label{lengthscale}
\end{equation}
We are looking for a class of models for which the ratio of the energy densities of CDM  and DE obeys a power law with respect to this length scale,
\begin{equation}
r=\dfrac{\bar{\rho}_{c}}{\bar{\rho}_{\Lambda}}=r_{0}l^{-\xi}\quad\Rightarrow\quad\dfrac{\dot{r}}{r}=-\dfrac{\xi}{3}\Theta.
\label{scalinggeneral}
\end{equation}
These models are characterized by the free parameter $\xi$. Note that in a homogeneous and isotropic Friedmann-Lema\^{\i}tre-Robertson-Walker (FLRW) space-time
the expansion scalar is equal to three times the Hubble parameter and the length scale $l$ coincides with the scale factor $a$ of the RW metric. Relation (\ref{scalinggeneral}) covariantly generalizes the model by Dalal et al. \cite{dalal} which was restricted to the background dynamics.
Different properties of this model and its observational consequences
have
been investigated in \cite{ZPGRG,somasri,alcaniz,david}.
Based on the covariant generalization (\ref{scalinggeneral}) we are able to study the perturbation dynamics of the scaling model.
The $\Lambda$CDM model is recovered as the special case $\Theta =\frac{3}{a}\mathcal{H}$, $w_{\Lambda}=-1$, and $\xi=3$.

\subsection{Background equations}

The \textit{ansatz} (\ref{scalinggeneral}) induces an interaction between CDM and DE. According to (\ref{bgenergy1}) with (\ref{q1dark}), their background energy balances take the form
\begin{eqnarray}
\bar{\rho}^{\prime}_{c}+3\mathcal{H}\bar{\rho}_{c}&=&-a\bar{Q},\label{bgscalingcdm}\\
\bar{\rho}^{\prime}_{\Lambda}+3\mathcal{H}\left(1+w_{\Lambda}\right)\bar{\rho}_{\Lambda}&=&a\bar{Q}.\label{bgscalingde}
\end{eqnarray}
Combining the equations (\ref{scalinggeneral}), (\ref{bgscalingcdm}) and (\ref{bgscalingde}) and solving for the background interaction term yields
\begin{equation}
\bar{Q}=-3\mathcal{H}\bar{\rho}_{c}\bar{\rho}_{\Lambda}\dfrac{\xi/3+w_{\Lambda}}{a\left(\bar{\rho}_{c}
+\bar{\rho}_{\Lambda}\right)}.
\label{bgscalingq0}
\end{equation}
A non-linear interaction of this type induces a scaling (\ref{scalinggeneral}) of the ratio of the energy densities.
This coupling is  proportional to the product of the energy densities of the interacting components which we consider
to be more ``realistic" than most of the other interacting models with interactions just linear in the energy density of the components. It is only in the limit $\bar{\rho}_{(c)} \gg \bar{\rho}_{(\Lambda)}$, i.e. at high redshift, that one approaches a linear dependence $\bar{Q} \propto -3\mathcal{H}\bar{\rho}_{(\Lambda)}/a$ which is the preferred model  in many studies in the field \cite{gawela,salvatelli,wangAbd1,wangAbd2} .
 Obviously, the interaction vanishes for $\xi/3+w_{\Lambda}=0$.

Integrating  equations (\ref{bgscalingcdm}) and (\ref{bgscalingde}), we obtain the analytic solutions
\begin{eqnarray}
&\bar{\rho}_{c}=\bar{\rho}_{c0}a^{-3}\left(\dfrac{\Omega_{c0}+\Omega_{\Lambda0}a^{\xi}}{\Omega_{c0}+\Omega_{\Lambda0}}\right)
^{-1-3w_{\Lambda}/\xi}, \label{rhoscalingcdm}\\
&\bar{\rho}_{\Lambda}=\bar{\rho}_{\Lambda0}a^{-3+\xi}\left(\dfrac{\Omega_{c0}+\Omega_{\Lambda0}a^{\xi}}{\Omega_{c0}+\Omega_{\Lambda0}}\right)
^{-1-3w_{\Lambda}/\xi}.\label{rhoscalingde}
\end{eqnarray}
The Friedmann equation (\ref{friedmann1}) then provides us with an  expression for the Hubble rate,
\begin{equation}
\mathcal{H}^{2}=\dfrac{8\pi G}{3}a^{2}\left[\frac{\Omega_{\Lambda0}+\Omega_{c0}}{a^{3\left(1+w_{\Lambda}\right)}}
\left(\dfrac{\Omega_{\Lambda0}
+\Omega_{c0}a^{-\xi}}{\Omega_{\Lambda0}+\Omega_{c0}}\right)^{-3w_{\Lambda}/\xi}
+ \frac{\Omega_{b0}}{a^{3}}+\frac{\Omega_{r0}}{a^{4}}\right].
\label{scalinghubble}
\end{equation}
The present value $\mathcal{H}_{0}$ of the Hubble rate is, as usual, conveniently parametrized by
$\mathcal{H}_{0}=H_{0}=100h\ \mathrm{kms^{-1}Mpc^{-1}}$.

With (\ref{scalinghubble}) the background dynamics is explicitly known. Obviously, the existence of an analytic expression for the background Hubble rate will be useful in dealing with the perturbation dynamics.

\subsection{Perturbations equations}

To solve the first-order dynamics, an explicit expression for the perturbed interaction is required. As already mentioned, the spatial term has the
model independent form  (\ref{hatf}). For the temporal term $\hat{Q}$ we assume that the background relation (\ref{bgscalingq0})
is valid in general, thus on the perturbed level,
\begin{equation}
\hat{Q}=\left(\dfrac{\bar{\rho}_{c}\bar{\rho}_{\Lambda}}{\bar{\rho}_{c}+\bar{\rho}_{\Lambda}}\right)\left(w_{\Lambda}+\xi/3\right)
\left[\hat{\Theta}+\dfrac{3\mathcal{H}}{a\left(\bar{\rho}_{c}+\bar{\rho}_{\Lambda}\right)}
\left(\bar{\rho}_{c}\delta_{\Lambda}+\bar{\rho}_{\Lambda}\delta_{c}\right)\right],
\label{hqscaling}
\end{equation}
where, following \cite{jhumberto}, the first-order expansion scalar is,
\begin{equation}
\hat{\Theta} = -\frac{1}{a}\left(k^{2}v+ \frac{h^{\prime}}{2}\right).
\label{htheta}
\end{equation}
The complete set of equations for the dark-sector perturbations is obtained by inserting (\ref{hqscaling}) for $\hat{Q}$ into the first-order balance equations. For the case
$w_{\Lambda}=-1$ we combine (\ref{hqscaling}) and (\ref{htheta}) with the equations (\ref{deenergypert1}), (\ref{cdmenergypert})
and (\ref{cdmmomentumpert}),  for the case $w_{\Lambda}\neq -1$ the expressions (\ref{hqscaling}) and (\ref{htheta}) are combined with equations (\ref{deenergypert}),
(\ref{demomentumpert1}), (\ref{cdmenergypert}) and (\ref{cdmmomentumpert1}).

Many papers in the literature neglect the perturbations of DE and of the interaction term, taking $\delta_{(\Lambda)}=\hat{f}=\hat{Q}=0$. Here we take
into account these perturbations and we shall quantify their impact on the CMB power spectrum within a statistical analysis.

\section{Numerical analysis}
\label{numeric}

The formalism described above can be used now to test the scaling cosmology model against the observational data. The main motivation of this
chapter is to compare the parameter selection obtained by  background tests using SNe Ia data from the JLA sample \cite{JLA} and
baryon acoustic oscillations (BAO) data from 6dFGS\cite{bao1}, SDSS \cite{bao2}, BOSS CMASS \cite{bao3} and the WiggleZ survey \cite{bao4},
with a perturbative test using the
CMB power spectrum data from Planck \cite{planck}.
In the simplified analysis of this paper we focus on the impact of variations of the parameter $\xi$ on the spectrum.
Because of computational limitations, we restricted our analysis to the case $w_{\Lambda}=-1$.
We shall include, however, a qualitative discussion of how the spectrum changes for $w_{\Lambda}\neq -1$.
The SNe Ia and BAO analysis are based on
\begin{equation}
\chi^{2}(\theta)=\Delta y (\theta)^T\mathbf{C}^{-1} \Delta y(\theta)\,,
\label{chi2}
\end{equation}
where $\Delta y(\theta) = y_i-y(x_i;\theta)$ and $\theta$ are  the free parameters. The  $y(x_{i}\vert\theta)$  represent  the theoretical predictions for a given set
of parameters and $\mathbf{C}$  is the covariance matrix that in the case of JLA  is given in \cite{JLA}.
In the case of the BAO analysis,  6dFGS, SDSS and BOSS CMASS data are mutually uncorrelated and they are also not correlated
with WiggleZ data. However, we must take into account  correlations beetween WiggleZ data points  given in \cite{bao4}.
A proper CMB analysis using the  Planck likelihood of \cite{likelihood} will be provided elsewhere. Here, we tentatively ignore all the details of the Planck likelihood and rely on a simple $\chi^{2}$ analysis
in order to get an idea of how a varying $\xi$ influences the spectrum.
Since we are modifying only the dark sector, we have $\theta=\left\lbrace\Omega_{c0},h,\xi\right\rbrace$.
Moreover,
in the CMB spectrum analysis
we preliminarily fix the set of parameters $\left\lbrace\Omega_{b0},\tau,A_{s},n_{s}\right\rbrace$, where
$\tau$ is the optical depth, $A_{s}$ is the initial perturbation amplitude and $n_{s}$ is the spectral index,
using the best-fit values
for the TT+low P+lensing results of the $\Lambda$CDM model  in \cite{planck}.
In a more complete future study these parameters, in particular $A_{s}$ and $n_{s}$, will have to be considered as free parameters as well.
In the entire analysis, the Hubble rate $h$ is left free but at the end it is
marginalized in order to obtain $\Omega_{c0}-\xi$ contour curves.

\subsection{SNe Ia}

As is well known, SNe Ia tests are using the luminosity distance modulus
\begin{equation}
\mu=5\log\left[d_{L}\left(z\right)\right]+\mu_{0}
\label{muSN}
\end{equation}
with $\mu_{0}=42.384-5\log\left(h\right)$, where $d_{L}$ is the luminosity distance
\begin{equation}
d_{L}\left(z\right)=\left(z+1\right)H_{0}\int_{0}^{z}\dfrac{d\tilde{z}}{H\left(\tilde{z}\right)}.
\end{equation}
Here we extend a previous analysis \cite{alonso} by including a separately conserved baryon component
using the Hubble parameter (\ref{scalinghubble}).
This confirms that the SNe Ia analysis does not satisfactorily constrain the interaction parameter $\xi$, i.e., the luminosity distance modulus is not very sensitive to this
parameter. The results of our analysis are shown in FIG.~1 (continuous curves).

\begin{figure}[h]\label{fig1}
\subfigure{
\includegraphics[scale=0.6]{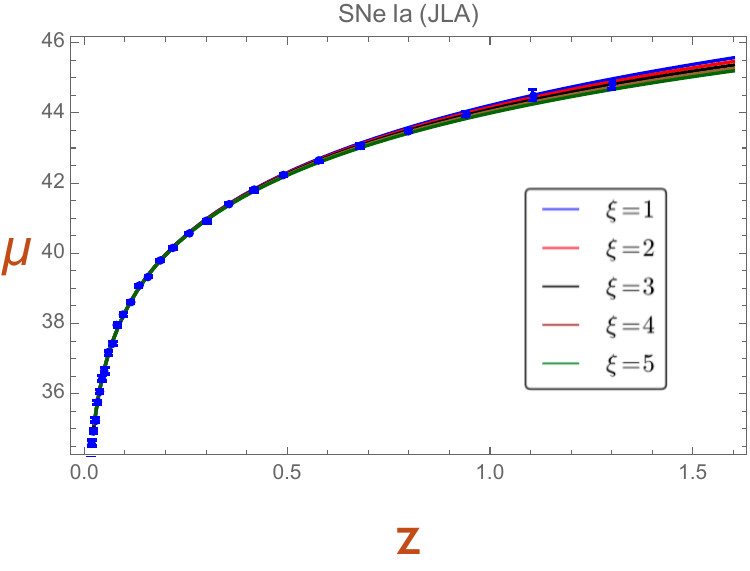}}
\subfigure{
\includegraphics[scale=0.5]{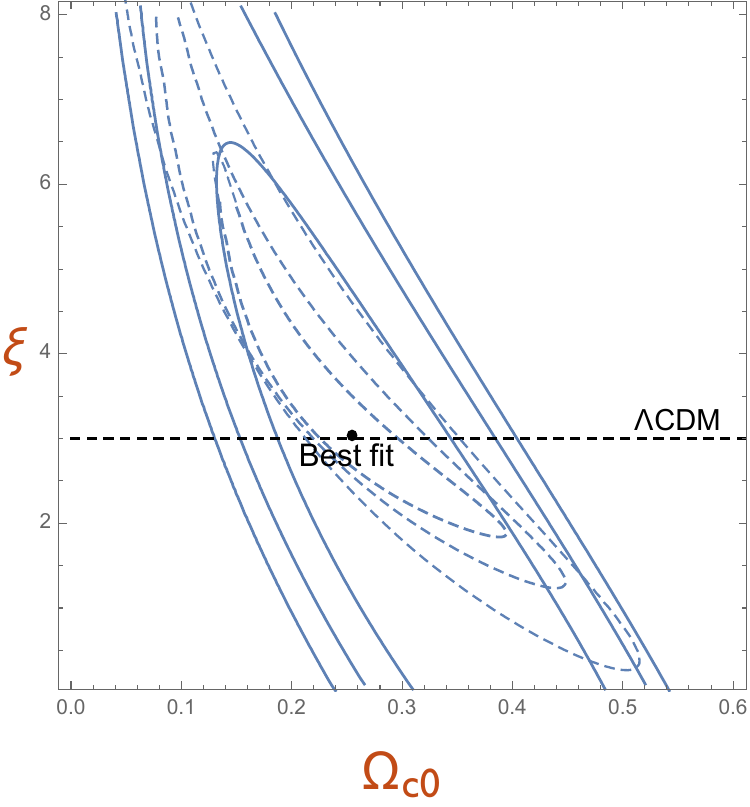}}
\caption{Left panel:
luminosity distance modulus for different values of $\xi$ with best-fit values for $\Omega_{c0}$ and $h$.
The curves demonstrate that the SNe Ia test is not well suited to constrain the parameter $\xi$.
Right panel: continuous contour lines (1$\sigma$, 2$\sigma$ and 3$\sigma$) resulting from the SNe Ia analysis
and dashed contour lines (1$\sigma$, 2$\sigma$ and 3$\sigma$) resulting from the BAO analysis, both with
three free parameters ($\Omega_{c0}$, $\xi$ and $h$). Here we marginalized over $h$. The best-fit values
of the joint analysis are $\Omega_{c0}=0.261\pm^{0.028}_{0.029}$
and $\xi=3.02\pm^{0.39}_{0.40}$ (1$\sigma$).}
\end{figure}

\subsection{BAO}

Our BAO-data analysis is based on the estimator $r_s(z_d)/D_{V}(z)$. In oder to construct
its theoretical counterpart we compute the dilation scale $D_V(z)$  \cite{Eiseinstein2005}
\begin{equation}
D_{V}(z)=\left[(1+z)^{2}D^{2}_A(z)\frac{cz}{H(z)}\right]^{1/3}\,,
\end{equation}
where $D_A(z)=(1+z)^{-1}r(z)$ is the angular diameter distance,
$r(z)$ is the comoving distance to the redshift $z$
\begin{equation}
r(z)=\int^z_0 \dfrac{d\tilde{z}}{H\left(\tilde{z}\right)}
\end{equation}
and $r_s(z_d)$ is the comoving sound horizon at the baryon-drag epoch $z_d$.
\begin{equation}
r_s(z_d)=c\int^\infty_{z_d} \dfrac{ c_s(\tilde{z})}{H\left(\tilde{z}\right)} d\tilde{z} \qquad  c_s(z)=\frac{1}{\sqrt{3\left(1+\frac{3\Omega_{b0}}{4 \Omega_{r0}}(1+z)^{-1}\right)}} \,.
\end{equation}
The results of the BAO analysis are shown in the right panel of FIG.~1 (dashed curves).
We  observe that also BAO tests constrain the interaction parameter $\xi$ weakly. A joint test SNIa+BAO was also  performed
and results are shown in FIG.~3 (dashed curves).

\subsection{CMB}

In order to obtain the CMB power spectrum for our model, we modify the
CAMB code by implementing the perturbation  equations (\ref{deenergypert1}), (\ref{cdmenergypert}),
(\ref{cdmmomentumpert}), (\ref{hqscaling}) and (\ref{htheta}).
To integrate the equations, the initial conditions were fixed following
\cite{Ballesteros}, which are valid in the non-adiabatic case and for more general sound speeds.
It turns out that the CMB spectrum is much more sensitive
to the interaction parameter than the SNe Ia  and BAO data.

Using CAMB recursively, we found the values for $\Omega_{c0}$ and $\xi$ that best fit the \textit{Planck} data.
The values for $\xi$ are considerably stronger constrained than by the JLA and BAO data.
Small deviations from the $\Lambda$CDM model are still admitted which is in accordance with conclusions based on large-scale structure data \cite{alonso}.

The upper panel of FIG.~2 shows the CMB spectrum for different values of the interaction parameter $\xi$.
For values of $\xi$ substantially different from the $\Lambda$CDM value $\xi = 3$
the CMB spectrum completely disagrees with the data even though the position of the first peak may be correct. The lower
panel of FIG.~2 shows the contour curves of the $\xi$-$\Omega_{c0}$ plane based on the \textit{Planck}
data and a summary of our studies is presented in Table~1. The degeneracy of  the parameter $\xi$ in the SNe Ia and BAO  analysis
is largely broken by the CMB analysis. This becomes very evident in the superposition of contour curves in FIG.~3.
The changes in the CMB spectrum due to differences in the parameter $\xi$ may be compared with the consequences of other parameter changes (see \cite{tegmark}).
On the largest scales, the ISW plateau is modified, an effect that is also caused by a variation of
$\Omega_{\Lambda 0}$ in the standard model. On smaller scales, a varying $\Omega_{\Lambda 0}$ changes the positions of the peaks but not their heights. On the other hand, the heights vary if the standard-model $\Omega_{c 0}$ is changed.
So, the variation of the height due different $\xi$ seems to indicate that there remains a degeneracy between $\Omega_{c 0}$ and $\xi$.
A variation in the baryon fraction changes the heights in the opposite direction as the $\Omega_{c 0}$ variation does.
Focusing on the first peak, there seems to be a similarity between changes in the baryon fraction (see \cite{luciano}) and changes in $\xi$. Increasing values of $\xi$ influence the height in a similar way as decreasing baryon fractions do. But for higher peaks this analogy does no longer hold.
Variations in $\xi$ do not result in changes of the types that are induced by variations of $A_{s}$ or $n_{s}$.
A variation of the spectral index rotates the CMB spectrum and a variation of the scalar amplitude increases or decreases the height of all peaks simultaneously \cite{tegmark}. None of these effects is simulated by variations in $\xi$. This seems to indicate that our fixing of $A_{s}$ and $n_{s}$ in this preliminary analysis is not inconsistent. A movie that visualizes the impact of $\xi$ on the CMB spectrum can be found under \cite{luciano2}.

\begin{figure}[h]
\subfigure{
\includegraphics[scale=0.7]{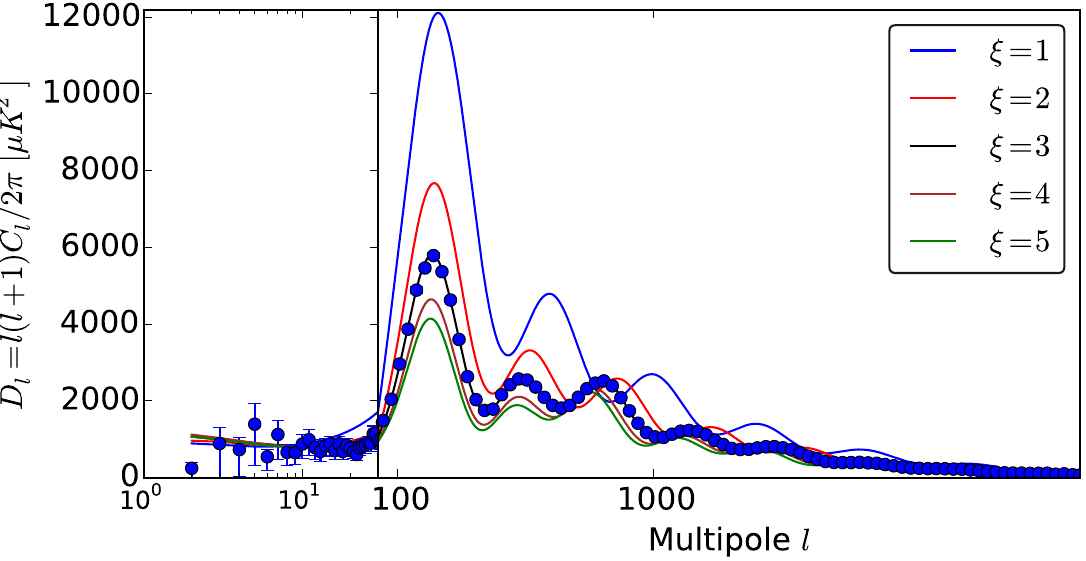}}\n{fig2a}\\
\subfigure{
\includegraphics[scale=0.7]{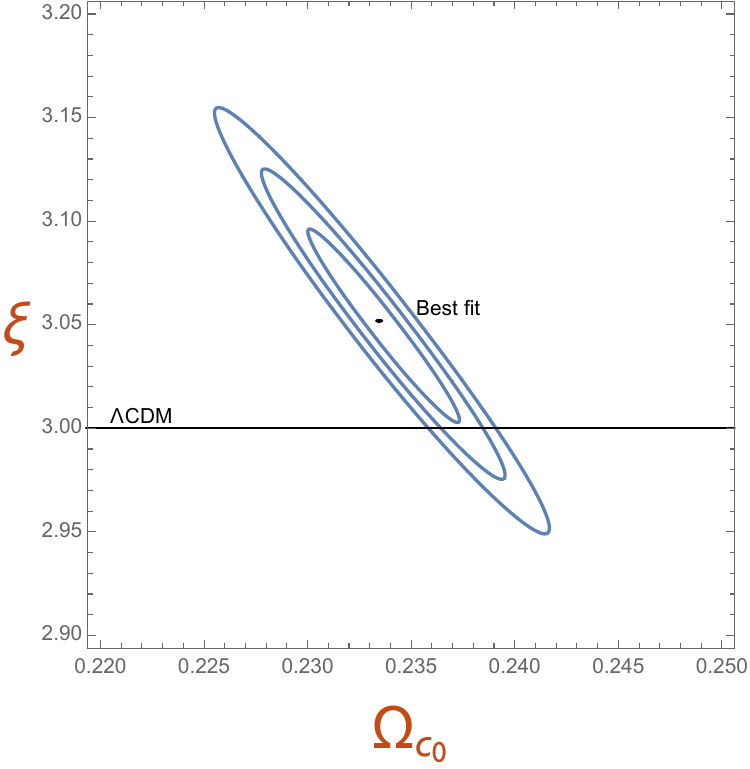}}\label{fig2b}
\caption{Top panel: CMB spectrum for different values of $\xi$ with best-fit values for $\Omega_{c0}$ and $h$.
The curves demonstrate that parameter values admitted by the SNIa and BAO analysis may be inconsistent with the CMB spectrum, even if the position of the first peak is approximately correct.
Bottom panel: contour lines (1$\sigma$, 2$\sigma$ and 3$\sigma$) resulting from the CMB analysis with three free parameters ($\Omega_{c0}$, $\xi$ and $h$). Here we marginalized over $h$. The best-fit values are $\Omega_{c0}=0.2334\pm^{0.0038}_{0.0033}$ and $\xi=3.052\pm^{0.047}_{0.048}$.}
\end{figure}

\begin{table}
\label{tab}
\begin{center}
\begin{tabular}{|c|c|c|c|}\hline
\multicolumn{4}{|c|}{Numerical results ($\pm 1\sigma$)} \\ \hline\hline
Test & $\Omega_{c0}$ & $h$ & $\xi$ \\ \hline
SNe Ia (JLA) + BAO & $0.261\pm^{0.028}_{0.029}$ & $0.701\pm^{0.003}_{0.004}$ & $3.02\pm^{0.39}_{0.40}$ \\\hline
CMB (Planck) & $0.2334\pm^{0.0038}_{0.0033}$ & $0.6901\pm^{0.0015}_{0.0019}$ & $3.052\pm^{0.047}_{0.048}$ \\\hline
\end{tabular}
\end{center}
\caption{Numerical results}
\end{table}

\begin{figure}
\label{fig3}
\begin{center}
\includegraphics[width=0.65\textwidth]{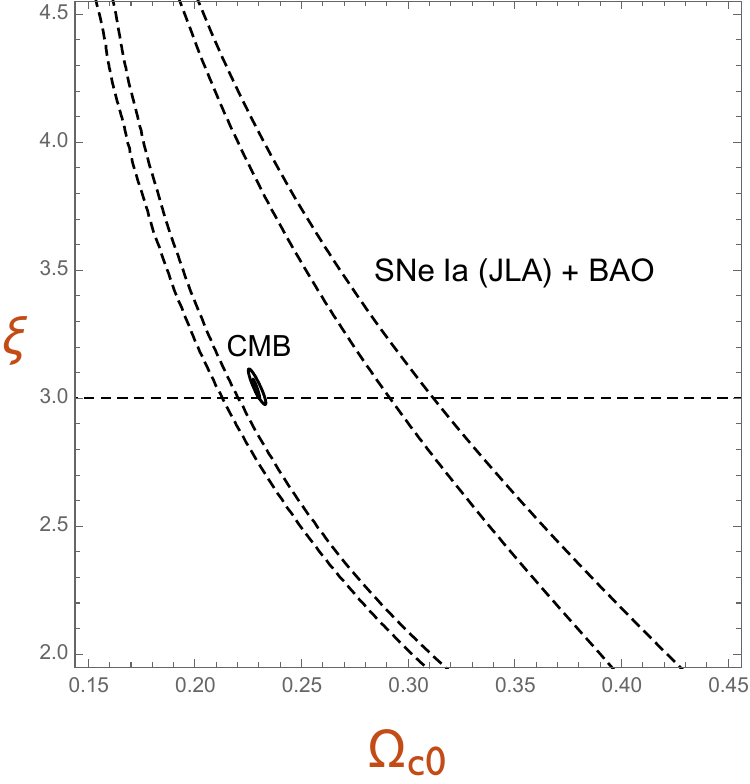}
\label{}	
\caption{Contour lines (1$\sigma$ and 2$\sigma$) for the combined CMB and SNIa+BAO analysis. Use of the CMB data drastically reduces the degeneracy in the parameter $\xi$.}
\end{center}
\end{figure}

\subsection{Role of DE perturbations}

Different from the $\Lambda$CDM model, any dynamical DE model is necessarily equipped with inhomogeneities not only of the matter distribution but also of the DE component itself.
Our basic set of perturbation equations allows us to quantify the potential relevance of these DE perturbations.
To this purpose we compare the CMB power spectrum for the best fit values of
$\Omega_{c0}$, $h$ and $\xi$ for the general case with $\delta_{\Lambda}\neq 0$, $\hat{Q}\neq 0$ and $\hat{f}\neq 0$ (our analysis so far) with a simplified version of the model with an interaction only in the background, assuming
vanishing DE perturbations, i.e., $\delta_{\Lambda}=\hat{Q}=\hat{f}=0$.
As FIG.~4 shows, the power spectra for both cases differ only very slightly at very large scales.
On smaller scales which here also include the scale of the first acoustic peak, we have identical results with and without DE perturbations.
This means, apart from the integrated Sachs-Wolfe effect, the essential features of the CMB remain unaffected.
For practical purposes this may justify the use of a model with vanishing DE perturbations.
The point is, however, that such a statement is possible only after the corresponding calculation has been done.

In FIG.~5 we show how the spectrum for  $\xi = 3$  changes if we deviate from $w_{\Lambda}=-1$. For a better visualization we used the drastically different values  $w_{\Lambda}=-0.5$ and $w_{\Lambda}=-1.5$. The position of the first peak does not change much, but the height does. The higher peaks change both in height and position. Combinations of $w_{\Lambda}=-0.5$ and $w_{\Lambda}=-1.5$ with  $\xi = 2.5$ and
$\xi = 3.5$ lead to qualitatively similar results.

\begin{figure}
\label{fig4}
\begin{center}
\includegraphics[width=0.9\textwidth]{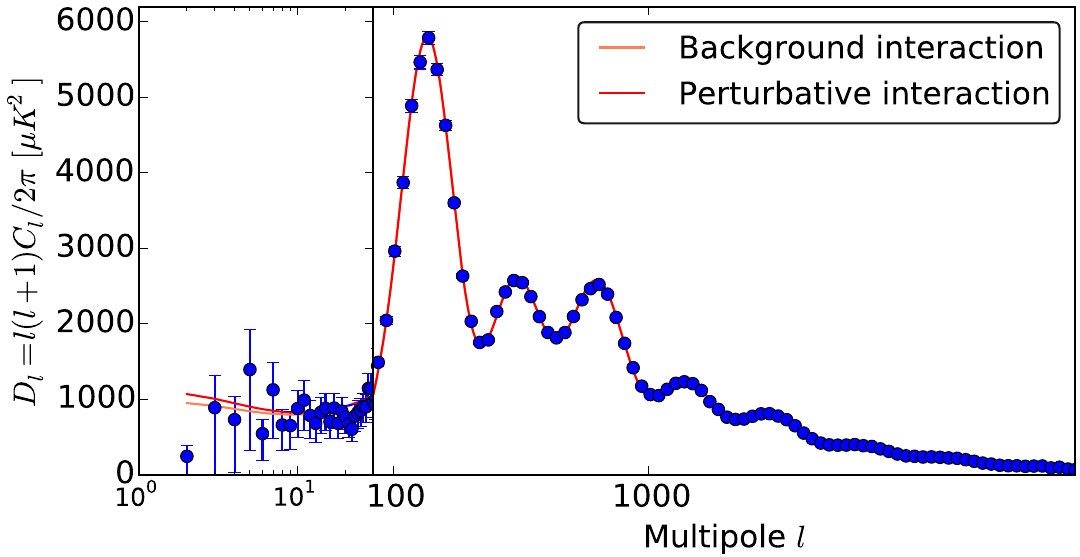}
\label{}	
\caption{CMB spectrum for the analysis with three free parameters ($\Omega_{c0}$, $\xi$ and $h$).
The best-fit values are $\Omega_{c0}=0.2334$, $\xi=3.052$ and $h=0.6901$. The case where the interaction is only taken into account in the background equations and $\delta_{\Lambda}=\hat{Q}=\hat{f}=0$ is almost identical with the general case with $\delta_{\Lambda}\neq 0$, $\hat{Q}\neq 0$ and $\hat{f}\neq 0$ except at the largest scales. }
\end{center}
\end{figure}

\begin{figure}
\label{fig5}
\begin{center}
\includegraphics[width=0.9\textwidth]{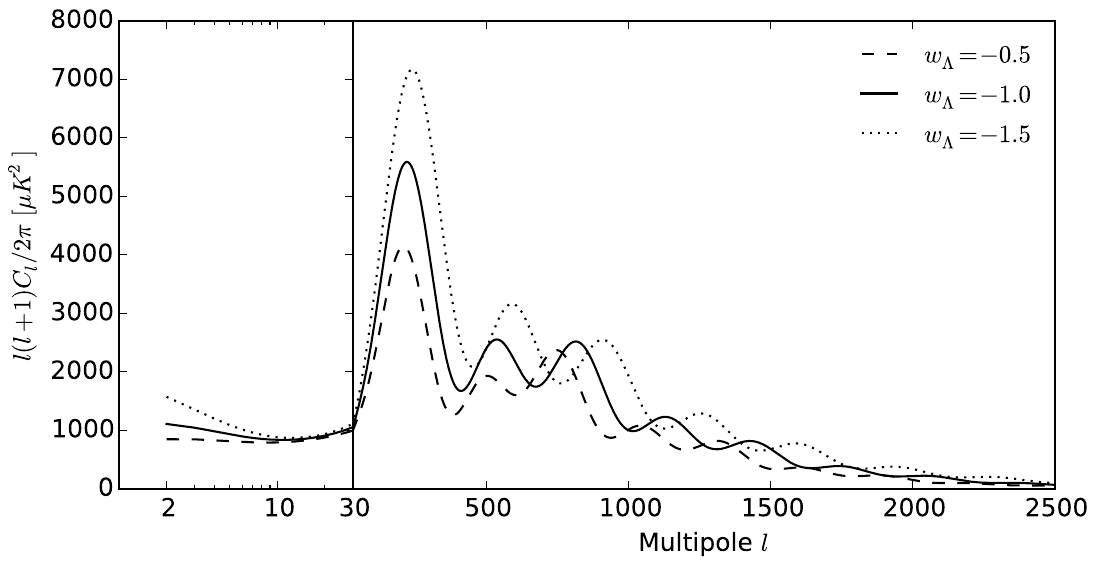}
\label{}	
\caption{Change of the CMB spectrum for $\xi=3$ if $w_{\Lambda}\neq-1$.  For $w_{\Lambda}>-1$ the height of the first peak is reduced, for $w_{\Lambda}<-1$ it is enhanced. The position of the first peak is almost unchanged. The higher peaks do change both in height and position. }
\end{center}
\end{figure}


\subsection{Matter power spectrum}

Our analysis of the CMB perturbations constrains $\xi$ to stay close to the $\Lambda$CDM value $\xi=3$.
This complements the results of an earlier study based on large-scale structure data \cite{alonso}.
Here we illustrate the change of the matter-power spectrum in dependence on $\xi$ and compare the results with those of the $\Lambda$CDM model.
FIG.~6 shows the dependence of the matter power spectrum on $\xi$ for $w_{\Lambda}=-1$.
Differences from $\xi=3$ result in increasing or decreasing powers on large scales together with corresponding
distortions of the shape at BAO scales which do not match
the large-scale structure observations.
The calculated matter distribution is generally different from the observed galaxy distribution.
We leave a thorough statistical analysis of the matter power spectrum, including issues of bias, for future work.
Under \cite{luciano2} one finds a movie that shows the impact of variation in $\xi$ on the matter power spectrum.


\begin{figure}
\label{fig6}
\begin{center}
\includegraphics[width=0.65\textwidth]{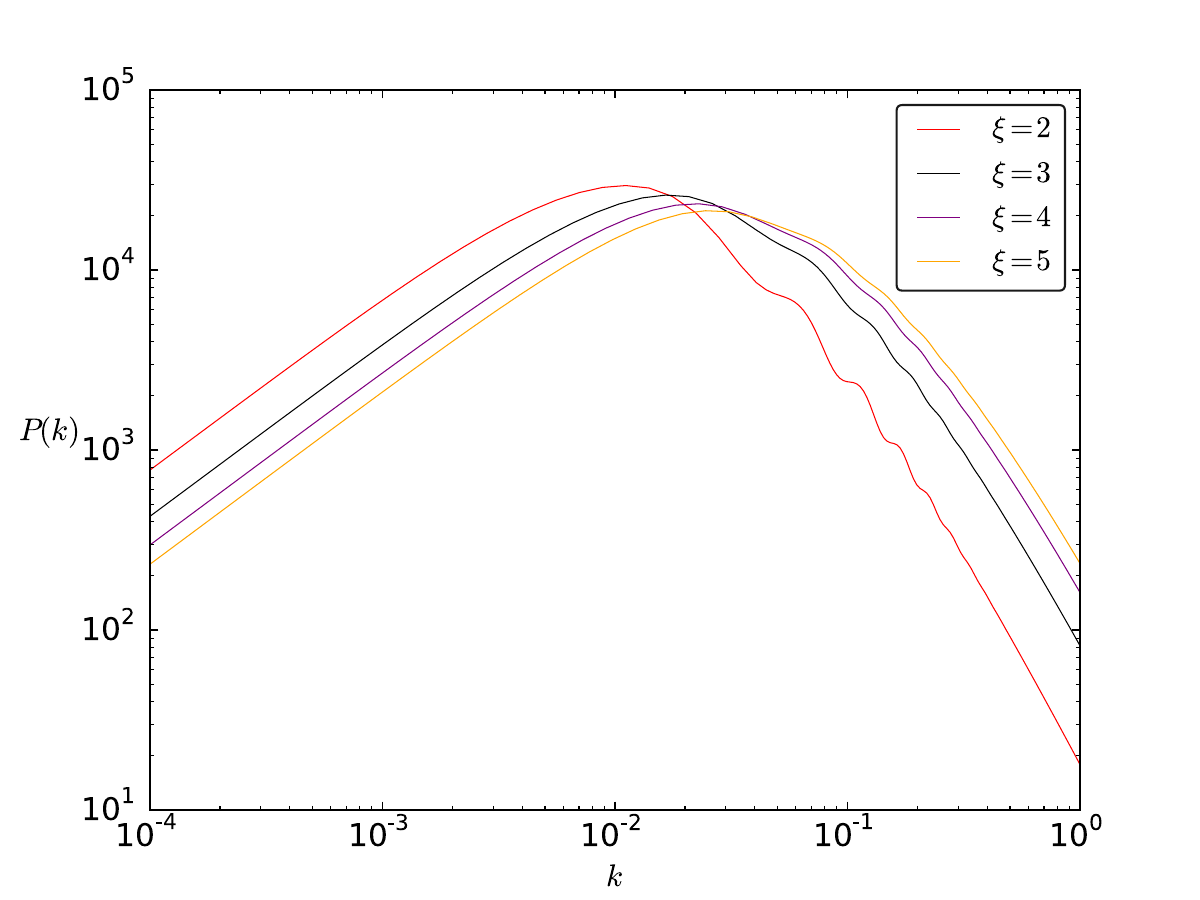}	
\caption{Matter power spectrum $P(k)$ for different values of $\xi$. The case $\xi=3$ corresponds to the
$\Lambda$CDM model.}
\end{center}
\end{figure}

\section{Conclusions}
\label{discussion}

Our main result is the calculation of the CMB spectrum in dependence of the model parameter $\xi$ where any value
$\xi \neq 3$ represents a deviation from the standard $\Lambda$CDM model due to a non-linear, non-gravitational interaction between DM and DE with  $w_{\Lambda}=-1$.
The details of the result are preliminary in the sense that we also made use here of some of the best-fit values for the $\Lambda$CDM model itself.
Our study can be seen as a first step towards a more accurate future analysis, avoiding ``contamination" by the $\Lambda$CDM model.
On the basis of the Planck data we find the best-fit values $\xi=3.052$  and $\Omega_{c0}=0.2334$  which coincide with the standard model  at the 2$\sigma$ confidence level.
The CMB data allow us to constrain the parameter $\xi$ with considerably higher precision  than
the SNIa data of the JLA sample	and BAO data points do.
In this context we clarified that the mere position of the first peak is not a suitable criterion for assessing a model.
We also included a qualitative discussion of the change of the spectrum for  $w_{\Lambda}\neq -1$.
Perturbations of the DE component where shown to be negligible, except, perhaps, on extremely large scales.
Furthermore, we illustrated the influence of variations in $\xi$ on the matter power spectrum.
An advanced more complete analysis which is independent of standard-model results will be the subject of future work.
\\

\noindent
{\bf Acknowledgement:} Financial support by  CAPES, FAPES and CNPq (Brazil) is gratefully acknowledged.
WSHR was supported by FAPES (BPC No476/2013)  at the begining of this work.
This work has made use of the computing facilities of National Center
for Supercomputing (CESUP/UFRGS), and of the Laboratory of
Astroinformatics (IAG/USP, NAT/Unicsul), whose purchase was made
possible by the Brazilian agency FAPESP (2009/54006-4) and the INCT-A.

\end{document}